\documentstyle[11pt,mrs2001,epsfig]{article}

\begin{document}
\title{CLUSTERING IN THE LAS CAMPANAS DISTANT
       CLUSTER SURVEY}

\author{Anthony H. Gonzalez $^1$, Dennis Zaritsky $^2$, and Risa H. Wechsler $^3$}
\affil{$^1$Harvard-Smithsonian Center for Astrophysics, 60 Garden Street, 
       Cambridge, MA 02138}
\affil{$^2$Steward Observatory, University of Arizona, 933 North Cherry Avenue,
           Tuscon, AZ 85721}  
\affil{$^3$Department of Physics, University of California, Santa Cruz, CA 95064}

\begin{abstract}
We utilize a sample of galaxy clusters at 0.35$<$$z$$<$0.6 drawn from
the Las Campanas Distant Cluster Survey (LCDCS) to provide the first
non-local constraint on the cluster-cluster spatial correlation
function. The LCDCS catalog, which covers an effective area of 69
square degrees, contains over 1000 cluster candidates. Estimates of
the redshift and velocity dispersion exist for all candidates, which
enables construction of statistically completed, volume-limited
subsamples. In this analysis we measure the angular correlation
function for four such subsamples at $z$$\sim$0.5. After correcting for
contamination, we then derive spatial correlation lengths via Limber
inversion. We find that the resulting correlation lengths depend upon
mass, as parameterized by the mean cluster separation, in a manner
that is consistent with both local observations and CDM predictions
for the clustering strength at $z$=0.5.
\end{abstract}

\section{Introduction}

The spatial correlation function of galaxy clusters provides an
important cosmological test, as both the amplitude of the correlation
function and its dependence upon mean intercluster separation are
determined by the underlying cosmological model. In hierarchical
models of structure formation, the spatial correlation length, $r_0$,
is predicted to be an increasing function of cluster mass, with the
precise value of $r_0$ and its mass dependence determined by $\sigma_8$
(or equivalently $\Omega_0$, using the constraint on
$\sigma_8-\Omega_0$ from the local cluster mass function) and the
shape parameter $\Gamma$.  Low density and low $\Gamma$ models
generally predict stronger clustering for a given mass and a greater
dependence of the correlation length upon cluster mass.     

In this paper we utilize the Las Campanas Distant Cluster Survey (LCDCS) to provide
a new, independent measurement of the dependence of the cluster
correlation length upon the mean intercluster separation ($d_c\equiv n^{1/3}$)
at mean separations comparable to 
existing Abell and APM studies. We first measure the angular
correlation function for a series of subsamples at $z\sim0.5$ and then
derive the corresponding $r_0$ values via the cosmological Limber
inversion \cite{efs91,hud96,peebles80}.  The resulting values
constitute the first measurements of the spatial correlation length
for clusters at $z\ga0.2$.  Popular structure formation models predict
only a small amount of evolution from $z=0.5$ to the present - a prediction
that we test by comparison of our results with local observations.

\section{The Survey}

The recently completed Las Campanas Distant Cluster Survey
is the largest published catalog of galaxy
clusters at $z\ga0.3$, containing 1073 candidates \cite{gon2001a}.  Clusters are detected in the
LCDCS as regions of excess surface brightness relative to the mean sky
level, a technique that permits wide-area coverage with a minimal
investment of telescope time.  The final statistical catalog covers an
effective area of 69 square degrees within a $78\deg\times1.6\deg$
strip of the southern sky ($860\times24.5$ $h^{-1}$ Mpc at $z$=0.5 for
$\Omega_0$=0.3 $\Lambda$CDM). Gonzalez et al. (2001$a$) also provide estimated
redshifts ($z_{est}$), based upon the brightest cluster galaxy (BCG)
magnitude-redshift relation, that are accurate to $\sim$15\% at
$z_{est}=0.5$, and demonstrate the existence of a correlation between
the peak surface brightness, $\Sigma$, and velocity dispersion,
$\sigma$. Together these two properties enable construction of
well-defined subsamples that can be compared directly with simulations
and observations of the local universe.                           

\subsection{The LCDCS Angular Correlation Function}

\begin{figure}
\plottwo{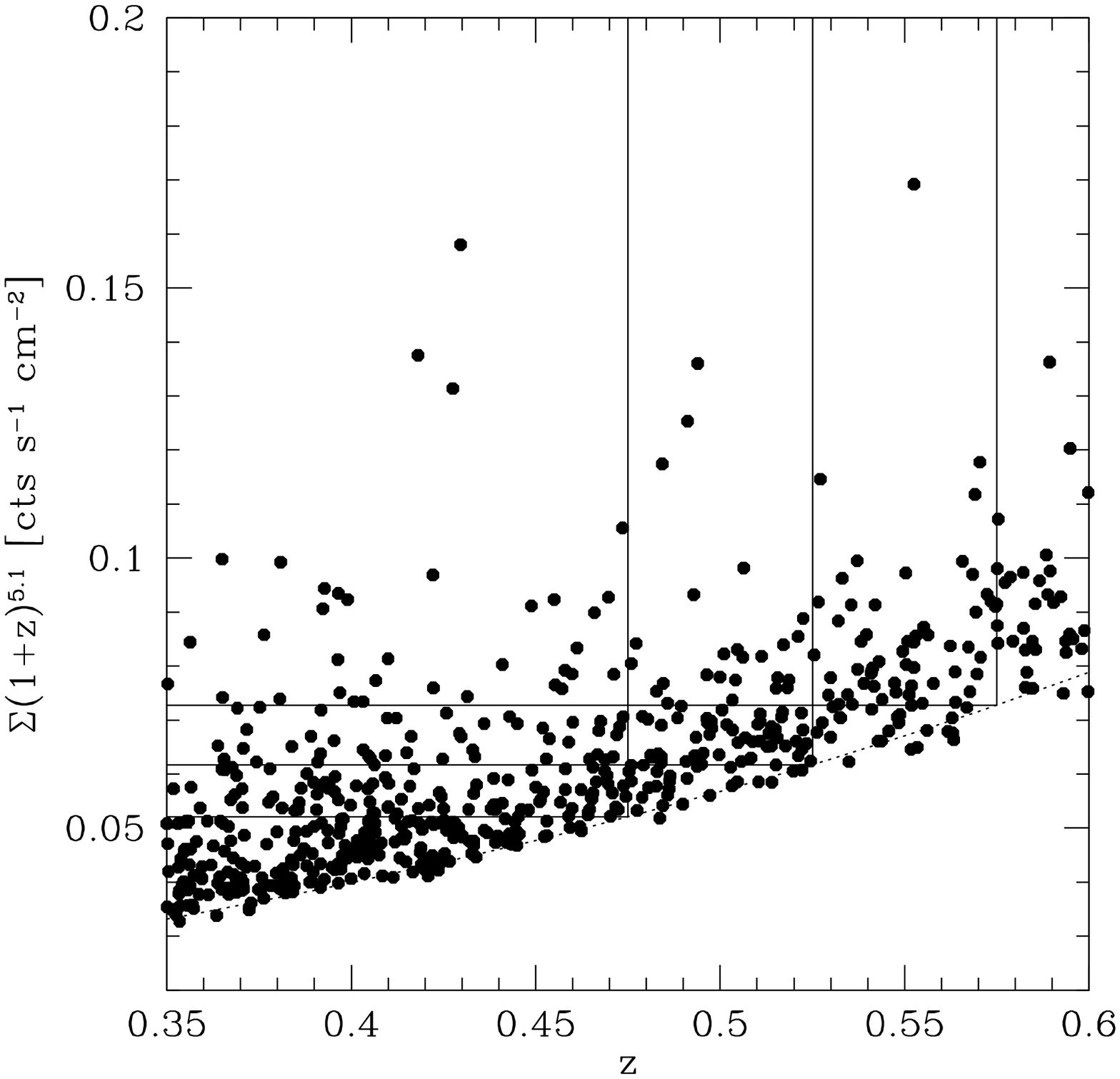}{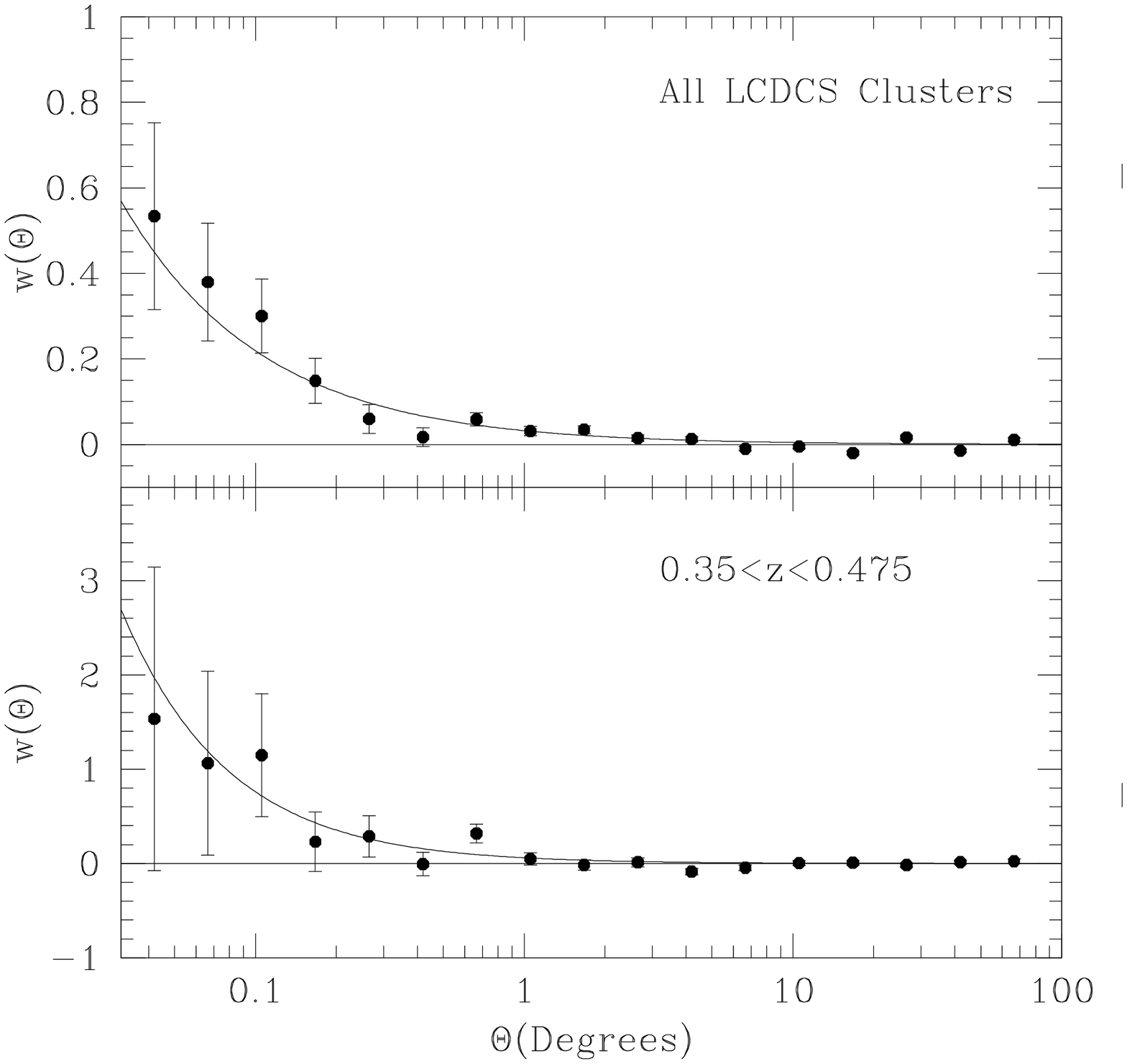}
\caption{ a/ Extinction-corrected surface brightness distribution of the LCDCS clusters 
used to construct subsamples. Gonzalez et al (20001$a$) found that
$\log \sigma \propto \log \Sigma (1+z)^5.1$, and so we employ this relation
to definty subsamples that are approximatlye dispersion-limited. The solid lines denote the redshift and
$\Sigma(1+z)^{5.1}$ thresholds of the subsamples. The dotted curve is
the surface brightness threshold of the full LCDCS catalog for E($\bv$)=0.05.
Errors in the estimated redshifts move individual data points along tracks 
similar to this curve;
b/ Angular correlation functions
for the entire LCDCS catalog and for a volume-limited subsample with
$0.35\le z<0.475$.  The solid lines are best-fit power law models.
\label{fig:fig1}}
\end{figure}

To compute the two-point angular correlation function, we use the 
estimator of Landy \& Szalay (1993). We measure the angular correlation function
both for the full LCDCS catalog and for three approximately velocity dispersion-limited
subsamples at $z$$\sim$0.5 (Figure \ref{fig:fig1}$a$).
We restrict all subsamples to $z_{est}$$>$0.35 to avoid incompleteness, while
the maximum redshift is determined by the surface brightness threshold of the
subsample. 

The angular correlation function for the entire LCDCS catalog is shown
in the upper panel of Figure \ref{fig:fig1}$b$, with logarithmic angular 
bins of width $\delta\log\theta$=0.2.  Modeling this correlation function 
as a power law, $ \omega(\theta)=A_\omega \theta^{1-\gamma}=
(\theta /\theta_0)^{1-\gamma},$
a least-squares fit for all LCDCS clusters over the range
2$\arcmin$-5$\deg$ yields $\gamma$=1.83$\pm$0.12 and
$\theta_0$=56$\pm22\arcsec$.
The angular correlation function for the lowest redshift subsample is
shown in the lower panel of Figure \ref{fig:fig1}$b$, overlaid with a best-fit power
law. 
We derive best-fit values both
allowing $\gamma$ to vary as a free parameter and also fixing
$\gamma$=2.1 --- equivalent to the best-fit value for the lowest
redshift subsample and the best fit value for the ROSAT All-Sky Survey
1 Bright Sample \cite{mos2000}.

We then apply a correction to these best-fit values to account for
the impact of false detections in the LCDCS catalog, which for this 
data set act to suppress the amplitude of the observed correlation
function.  If we assume that the contamination is spatially correlated
and can be described by a power law with the same slope as the cluster
angular correlation function (a reasonable approximation because for
galaxies -- which are likely the primary contaminant --
$\gamma$$\sim$1.8-1.9 [e.g. 2, 19]),
then the observed
value of $A_\omega$ is 
\begin{equation}
A_{ \omega}=A_{\rm cluster} (1-f)^2+A_{{\rm false}} f^2,
\end{equation}
where $f$ is the fractional contamination. For detections induced by
isolated galaxies of the same magnitude as BCG's at $z\sim0.35$ (and
identified as galaxies by the automated identification criteria
described in Gonzalez et al. (2001$a$), we measure that $A_{\rm gal}$ is
comparable to $A_{\omega_A}$, the net clustering amplitude for all
LCDCS candidates at 0.3$<$$z_{est}$$<$0.8. For detections identified as low
surface brightness galaxies (including some nearby dwarf galaxies) we
measure $A_{\rm LSB}\sim10A_{\omega_A}$. While these systems are
strongly clustered, we expect that they comprise less than half of the
contamination in the LCDCS.  For multiple sources of contamination the
effective clustering amplitude $A_{\rm false}=\sum A_i f_i^2/(\sum f_i)^2$, so
the effective clustering strength of the contamination is $A_{\rm false}\la
2.5 A_{\omega_A}$ even including the LSB's.           

\section{The Spatial Correlation Length}
\begin{figure}
\plottwo{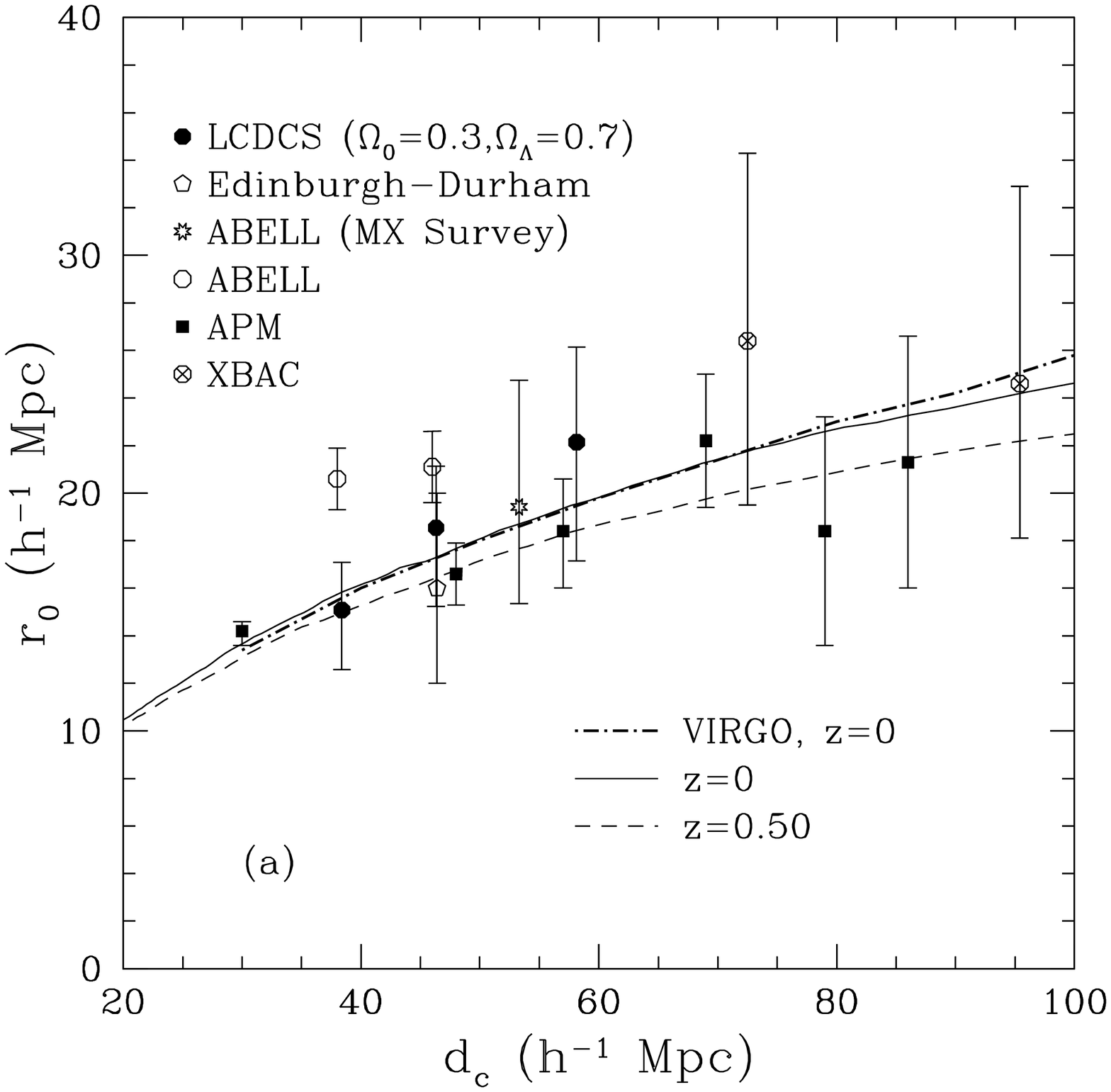}{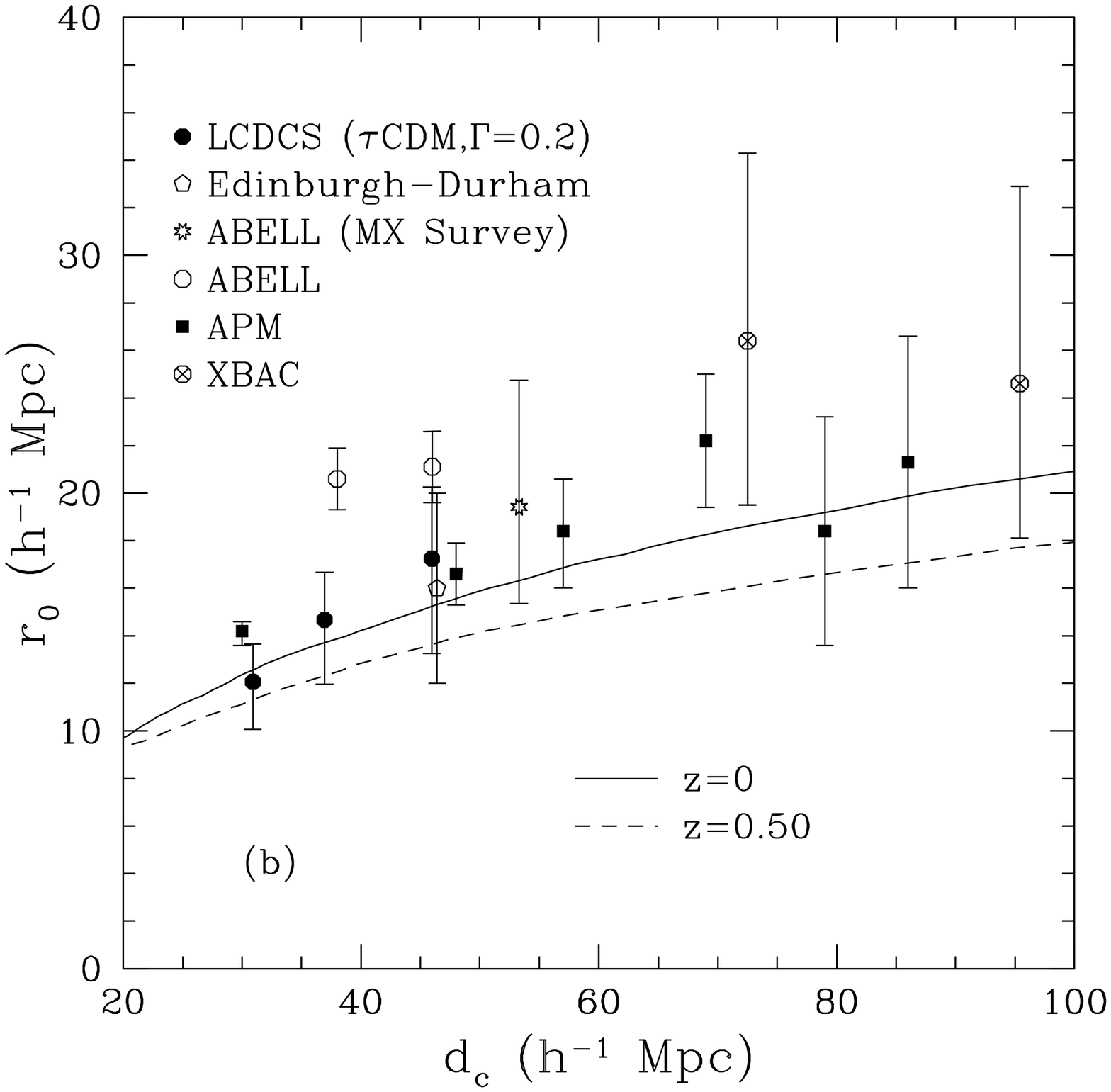}
\caption{ 
a/  Comparison of the LCDCS data with local samples and theoretical
predictions for $\Lambda$CDM ($\Omega$=0.3). The error bars 
on the LCDCS data correspond to the 1-$\sigma$ statistical uncertainty. 
Overlaid lines correspond to analytic predictions ($\Gamma$=0.2) for 
$z$=0 and $z$=0.5, and to results from the Virgo Consortium Hubble Volume simulations;
b/  Same plot for $\tau$CDM.
\label{fig:cfvslocal}} 
\end{figure}

The observed angular correlation function can be used to determine the
three-space correlation length if the redshift distribution of the
sample is known. This is accomplished via the cosmological Limber
inversion \cite{efs91,hud96,peebles80}. For a power-law correlation
function with redshift dependence $f(z)$, 
\begin{equation} 
       \xi(r)=\left(\frac{r}{r_0}\right)^{-\gamma}\times f(z).
\end{equation}        
The corresponding comoving spatial correlation length is $r_0(z)=r_0
f(z)^{1/\gamma}$, and the Limber equation is
\begin{equation}                    
r_o^\gamma=A_\omega\frac{c}{H_0}\frac{\Gamma(\gamma/2)}{\Gamma(1/2)\Gamma[(\gamma-1)/2]} 
           \left[ 
	   \frac{\int_{z1}^{z2} (dN/dz)^2 E(z) D_{A}(z)^{1-\gamma}
	   f(z) (1+z) dz}                                                      
	   {\left(\int_{z1}^{z2} (dN/dz) dz\right)^2}\right]^{-1},
\end{equation} 
where $dN/dz$ is the redshift distribution of the sample,
$D_{A}(z,\Omega_0, \Omega_\Lambda)$ is the angular diameter distance,
and $E(z)$ is defined as in Peebles (1993).
Because little evolution in the
clustering is expected over the redshift intervals spanned by our
subsamples (see Figure \ref{fig:cfvslocal}), $f(z)$
can 
be pulled out of the integral.

For the LCDCS we estimate the true redshift distributions of our
subsamples based upon the observed distribution of estimated
redshifts. If we approximate the redshift error distribution as
Gaussian with $\sigma_z/z\approx0.14$ at $z$=0.5 \cite{gon2001a},
then the actual redshift distribution $dN/dz$ is approximately equal
to the observed redshift distribution $dN_{obs}/dz$ for a given
subsample convolved with this Gaussian scatter. To test the validity
of this approach, we also try modeling $dN/dz$ using the theoretical
mass function of Sheth \& Tormen (1999) convolved with redshift uncertainty.
Comparing these two methods we find that the derived spatial
correlation lengths agree to better than 3\% for all
subsamples.\footnote{This result is insensitive to the exact mass
threshold assumed for the theoretical mass function.}

\begin{table} \begin{center}
\caption{Spatial Correlation Lengths}
\begin{tabular}{ccccccc}
\hline \hline 
{$z$} & \multicolumn{2}{c}{$\Lambda$CDM ($\Omega_0=0.3$) }  &
\multicolumn{2}{c}{OCDM ($\Omega_0=0.3$)} &
\multicolumn{2}{c}{$\tau$CDM($\Gamma$=0.2)}   \\    
range & $d_c$ & $r_0$ & $d_c$ & $r_0$ & $d_c$ & $r_0$ \\
\hline \\
0.35-0.475  & 38.4 & 15.1$^{+2.0}_{-2.5}$ & 33.8 & 13.2$^{+1.8}_{-2.2}$ & 30.9 & 12.1$^{+1.6}_{-2.0}$ \\
0.35-0.525  & 46.3 & 18.5$^{+2.6}_{-3.3}$ & 40.6 & 16.1$^{+2.3}_{-2.9}$ & 36.9 & 14.7$^{+2.0}_{-2.7}$ \\
0.35-0.575  & 58.1 & 22.1$^{+4.0}_{-5.0}$ & 50.8 & 19.1$^{+3.4}_{-4.3}$ & 46.0 & 17.3$^{+3.0}_{-4.0}$ \\  
\\
\hline \hline
\multicolumn{7}{l}{Note --- Units of $r_0$ and $d_c$ are $h^{-1}$ Mpc.}\\
\end{tabular} 
\end{center}
\label{tab:ro} \end{table}

\section{Results and Comparison with Local Data}
Table \ref{tab:ro} lists the correlation lengths ($r_0$) and mean intercluster
separations ($d_c$) that we derive for the
three LCDCS subsamples. The values of $r_0$ and 
$d_c$ are cosmology-dependent, so we list these quantities for three
different cosmologies --- $\Lambda$CDM ($\Omega_0$=0.3,$\Gamma$=0.2),
OCDM ($\Omega_0$=0.3,$\Gamma$=0.2), and $\tau$CDM ($\Gamma$=0.2). 
We opt to fix $\gamma$=2.1 when deriving the correlation lengths in 
Table \ref{tab:ro}
because $\gamma$ is strongly covariant with $A_\omega$ and
hence poorly constrained by the LCDCS data set. 
In no instance does this choice alter the derived $r_0$ value by more than 10\% from the
value obtained when $\gamma$ is treated as a free parameter, but we
caution that it  can systematically bias the observed dependence
of $r_0$ upon $d_c$.
For subsamples with best-fit
values of $\gamma$$>$2.1, fixing gamma slightly increases the derived
value of $r_0$, which results in a mild steepening of the dependence
of $r_0$ upon $d_c$ for the LCDCS subsamples.
                                                  
The cluster correlation function is predicted to not evolve
significantly between $z$=0.5 and the present; we thus compare the
LCDCS results directly with local observations, as is shown in Figure
\ref{fig:cfvslocal} for $\Lambda$CDM and $\tau$CDM.
For both cosmologies the LCDCS values of $r_0$ are
comparable to those from the Edinburgh-Durham Galaxy Catalogue
\cite{nic92}, APM survey \cite{cro97}, and the MX Survey northern sample
\cite{mil99}, but smaller than those found by
Peacock \& West (1992) for the Abell catalog. Also shown are the lowest $d_c$
data points for the XBAC catalog \cite{aba98}, which probe
higher masses than our study.
                                       
We also plot theoretical predictions
for comparison with the observational data. Results from the Virgo
Consortium Hubble Volume simulations are shown as a dot-dash line in
Figure \ref{fig:cfvslocal}$a$ \cite{col2000}.  The other lines
are analytic predictions based upon the work of Sheth \& Tormen (1999).
Like the APM data, the LCDCS results are consistent with the
low-density models (independent of $\Omega_\Lambda$). In contrast, the plotted $\tau$CDM model
systematically underestimates both the local and $z$=0.5
data. $\tau$CDM can be made to match the data only by decreasing $\Gamma$
to values that are inconsistent with constraints from the galaxy power
spectrum \cite{eis2000}.

To assess the robustness of these results, we also quantify 
potential systematic biases (see Gonzalez et al. 2001$b$
for more detail). We find that our results can only be significantly 
altered if the uncertainty in the estimated redshifts is underestimated.
If so, then the correlation lengths we derive would be systematically too small
(by $\sim$2 $h^{-1}$ Mpc if $\sigma_z/z$=0.25, for example). The second
most significant potential systematic arises from fixing $\gamma$, which as
mentioned above may lead us to overestimate the dependence of $r_0$ upon $d_c$.
Treating $\gamma$
as a free parameter would reduce the derived $r_0$ values for the
three subsamples (in order of increasing $d_c$) by 0\%, 7\%, and 10\%, but
would not qualitatively change our results.

\section{Discussion and Conclusions} 
The Las Campanas Distant Cluster Survey is the largest existing
catalog of clusters at $z$$>$0.3, providing a unique sample with which
to study the properties of the cluster population.  We use the
LCDCS to constrain the cluster-cluster angular correlation function,
providing the first measurements for a sample with a mean redshift
$z$$\ga$0.2.  From the observed angular correlation function, we
derive the spatial correlation length, $r_0$, as a function of mean
separation, $d_c$.  Only modest evolution in the clustering amplitude
is predicted between $z=0.5$ and present, and we thus compare our
results directly with local data.  We find that the LCDCS correlation
lengths agree with results from local samples, and observe
a dependence of $r_0$ upon $d_c$ that is comparable to the results of
Croft et al. (1997) for the APM catalog.  This clustering strength, its
dependence on number density, and its minimal redshift evolution are
consistent with analytic expectations for low density models, and with
results from the $\Lambda$CDM Hubble Volume simulations.
Consequently, while statistical uncertainty limits our ability to
discriminate between cosmological models, our results are in
concordance with the flat $\Lambda$CDM model favored by recent
supernovae and cosmic microwave background observations
\cite{boom01,pryke01,riess01}. 

A final result of this analysis is that it demonstrates the utility of
large catalogs like the LCDCS that are statistical in nature. While
the properties of any particular cluster in the LCDCS catalog are
rather uncertain, the properties of the sample as a whole are
well-defined, which is sufficient for constraining properties such as
the clustering strength and evolution in the comoving number density.
This statistical approach requires a relatively small investment in
telescope time, and so can be extended in the future to much larger
samples than the LCDCS.

\acknowledgements{ AHG acknowledges support from the 
Harvard-Smithsonian Center for Astrophysics.  DZ acknowledges financial 
support from NSF CAREER grant AST-9733111, and fellowships from the 
David and Lucile Packard Foundation and Alfred P. Sloan Foundation.  
RHW was supported by a GAANN fellowship at UCSC.}


\vfill
\end{document}